# Elastic interactions between colloidal microspheres and elongated convex and concave nanoprisms in nematic liquid crystals

Bohdan Senyuk[a] and Ivan I. Smalyukh[*abc]

[a]Department of Physics and Liquid Crystals Materials Research Center, University of Colorado at Boulder, Boulder, Colorado 80309, USA.
[b]Materials Science and Engineering Program, Department of Electrical, Computer, and Energy Engineering, University of Colorado at Boulder, Boulder, Colorado 80309, USA
[c]Renewable and Sustainable Energy Institute, National Renewable Energy Laboratory and University of Colorado at Boulder, Boulder, Colorado 80309, USA

*E-mail: ivan.smalyukh@colorado.edu

**Abstract**

We study mutual alignment and interactions between colloidal particles of dissimilar shapes and dimensions when dispersed in a nematic host fluid. Convex pentagonal and concave starfruit-shaped nanoprisms and microspheres induce dipolar or quadrupolar director structures. The ensuing elastic pair interactions between microspheres and nanoprisms are highly dependent on the nanoparticle shape being omnidirectionally attractive for convex prisms but strongly anisotropic for concave prisms. Elastic deformations due to spherical particles cause well-defined alignment of complex-shaped nanoparticles at distances much larger than the microsphere size. We characterize distance and angular dependencies of elastic pair interaction forces, torques, and binding energies. The studied elasticity-mediated self-assembly of metal and dielectric nanoparticles with dissimilar shapes and sizes opens new possibilities for self-assembly based fabrication of structured mesoscopic composites with predesigned properties.

# Introduction

Anisotropic interactions between colloidal particles in liquid crystals (LCs) attract a great deal of attention due to the possibility of their controlled orientated self-assembly into two- and three-dimensional structures useful for a broad range of applications.[1] These long-range colloidal interactions are mediated by particle-induced deformations of a LC director $n$ (a unit vector representing the average orientation of mesogenic molecules) that typically propagate to large distances and can be partially "shared" by colloids to minimize elastic energy.[2] The ensuing elasticity-mediated self-assembly of monodisperse microspheres,[1,3–20] spherical nanoparticles,[21,22] cylindrical colloids,[23–28] and particles with more exotic shapes[29–36] has been explored. However, possible applications of this type of colloidal self-assembly in fabrication of mesoscale metal–dielectric composites with novel properties will require the use of particles having dissimilar shapes, sizes, and chemical compositions.

Colloidal particles dispersed in LCs cause spatial deformations of a local director field $n(r)$, which depend on the confinement, size of colloids, as well as type and strength of anchoring at their surfaces.[1,2] For example, microspheres can induce both dipolar and quadrupolar configurations of $n(r)$ with accompanying singular point or line defects.[3,4] These induced distortions trigger elastic interactions between particles, which have been characterized for pairs of elastic dipoles,[5] quadrupoles,[6] dipole–quadrupole,[7,8] quadrupoles of different quadrupolar strength,[9,10] and mixed multipoles.[19,20] However, elastic interactions between colloids of highly dissimilar shapes, compositions, and dimensions (*e.g.* nanocolloids[21–23] and micro- and nanocolloids[36]) have rarely been explored.

Recently, using optical manipulation by laser tweezers, stable colloidal dimers were formed from dielectric microspheres and gold nanoprisms by optically forcing nanoprisms into defects accompanying the spheres.[36] These dimers require laser tweezers for assembly, although they are stable after turning off the tweezers. In this article, we explore colloidal self-assembly due to elasticity-mediated interactions between similar pentagonal gold nanoprisms and micron-sized colloids. We demonstrate very different elastic pair interactions of concave and convex pentagonal nanoprisms with microspheres. The ensuing well defined elasticity-mediated self-assembly of highly dissimilar particles may be important for facile large-scale fabrication of nanostructured composite materials with controlled properties.

## Materials and techniques

We have used a room-temperature nematic LC material pentyl cyanobiphenyl (5CB) from Frinton Laboratories, Inc. Elongated convex pentagonal (CPNs) and concave starfruit-like nanoprisms (CSNs) with length width dimensions of 800 × 150 nm and 500 × 100 nm, respectively, were obtained as ethanol-based dispersions from Nanopartz, Inc.[36] and are the same nanoparticles as studied in ref. 36. CPN and CSN both have $D_{5h}$ symmetry but pentagonal convex and star-like concave bases,[36] respectively (insets of Fig. 1a and d). Silica spherical particles (SPs) of 3 μm diameter (Duke Scientific) were treated with a surfactant [3-(trimethoxysilyl)propyl]octadecyl-dimethylammonium chloride (DMOAP) to induce a strong homeotropic (normal to the surface) alignment[11] of LC molecules (Fig. 1g). To obtain isolated well-separated colloids in a nematic host, particles were added to LC at low concentrations (<0.05 vol%). First, SPs were dispersed in LC by direct mixing and a 15–30 min sonication to break apart pre-existing aggregates. Then gold nanoparticles were dispersed from ethanol to toluene and added to the mixture of LC and microspheres; this was followed by evaporation of toluene. The resulting colloidal dispersion was sonicated for 30–60 min while slowly changing the temperature from an isotropic to the LC phase. LC-colloid dispersions were filled in between two glass substrates spaced by glass microfibers of 10 mm in diameter (setting a cell thickness $d$) and treated to induce homeotropic or planar (tangential to the surface) far-field alignment $\mathbf{n}_0$ of LC. To set surface boundary conditions at confining substrates, we dip-coated them in an aqueous (1 wt%) solution of DMOAP for homeotropic alignment and used unidirectionally rubbed thin films of spin-coated and crosslinked polyimide PI2555 (HD MicroSystem) for planar alignment.

An experimental setup built around an inverted Olympus microscope IX81 was used to perform dark-field, bright-field, polarizing and three-photon excitation fluorescence polarizing microscopy observations. We used a tunable (680–1080 nm) Ti:sapphire oscillator (140 fs, 80 MHz, Chameleon Ultra-II, Coherent) for three-photon self-fluorescence polarizing microscopy[37,38] (Fig. 1h). The excitation of 5CB molecules was performed at 870 nm and fluorescence was detected within a spectral range of 387–447 nm by a photomultiplier tube H5784-20 (Hamamatsu). The in-plane position of the focused excitation beam was controlled by a scanning unit (FV300, Olympus) and its polarization was varied using a half-wave plate mounted immediately before the 100× (numerical aperture NA = 1.4) oil objective. An Olympus 100× (NA = 0.6) oil objective and a dark-field condenser U-DCW (NA = 1.2) were used for the dark-field videomicroscopy observations. Motion of colloids was recorded with a CCD camera (Flea, PointGrey) at a rate of 15 fps. A holographic optical trapping system[12,37] operating at the wavelength λ = 1064 nm and integrated with the IX81 optical microscope was used for optical manipulations of particles.

## Results and discussion

### Shape-dependent nanoprism alignment and elastic distortions

Colloidal particles immersed into a uniformly aligned LC distort $n(r)$ and, if shape-anisotropic, align with respect to $n_0$ so that elastic and surface anchoring energies are minimized. Distortions of $n(r)$ around the particle can be characterized by a parameter[24,25] $\omega \sim R/l_a$, where $R$ is the radius of the particle and $l_a = K/W_a$ is the anchoring extrapolation length;[2] $K$ is the average elastic constant of LC and $W_a$ is the surface anchoring coefficient.[2] For typical LC materials, one finds $l_a \sim 0.1–5$ μm. SPs of radius $R_{SP} \approx 1.5$ μm with the strong homeotropic anchoring ($\omega \gg 1$) cause $n(r)$-deformations of dipolar symmetry with a point defect "hedgehog" next to SP[4,5] and an elastic dipole moment along $n_0$ (Fig. 1g); the $n(r)$-structure was verified by three-photon fluorescence imaging (Fig. 1h). In dark-field images, SP appears as a bright scattering ring on a dark background of LC with the hedgehog observed as a weakly scattering spot next to it (marked by a blue arrow in Fig. 1i).

Deformations of $n(r)$ due to nanoparticles are weak and short-ranged as $\omega \sim 0.05–1$. Dark-field (Fig. 1b, c and f) and polarizing (Fig. 1e) imaging reveals the orientation of CPNs and CSNs and the surrounding $n(r)$ (Fig. 1a and d), as described in ref. 36. CPNs align perpendicular to $n_0$ and are encircled by disclination loops of half-integer strength[4,13,36] (Fig. 1a–c); these nanoparticles freely rotate around $n_0$ (Fig. 1a–c). Since their lateral size is smaller than $l_a$, vertical surface boundary conditions partially relax at the sharp edges between side faces of these nanoprisms.[33] CPNs align with one of their side faces either parallel or perpendicular to $n_0$ (insets of Fig. 1a). Because of their pentagonal cross-sections, $n(r)$ distortions around CPNs are asymmetric with a broken symmetry plane either containing the half-integer disclination loop or being perpendicular to it (insets in Fig. 1a). This gives rise to an elastic dipole moment[19,20] $p$ always orthogonal to the elongated nanoprism and directed either parallel or perpendicular to $n_0$, as shown in the insets of Fig. 1a. CSNs orient along $n_0$ (Fig. 1d–f) and freely rotate around it while director bends around and follows the surface morphology of the nanoprism with the $D_{5h}$ symmetry. The ensuing $n(r)$ is of quadrupolar type,[4,8,10] with two surface point defects (called "boojums") at CSN ends (Fig. 1d).[36] The grooved surface relief of the concave CSN with $\omega \gtrsim 1$ (inset in Fig. 1d) forces $n(r)$ to follow the surface morphology of CSN instead of satisfying antagonistic surface boundary conditions exerted by its concave facetted faces. More detailed studies of this $n(r)$-structure are discussed in our previous work (ref. 36).

**Elastic pair interactions between spheres and convex prisms**

In a planar nematic cell, SP and CPN experience attraction from all studied initial positions (Fig. 2a and b) when placed by optical tweezers in a close vicinity of $(4-6)R_{SP}$ to each other. This omnidirectional attraction is largely due to the fact that SP creates a highly distorted $n(r)$ around it and displacing some of these distortions by a nanoparticle reduces the total elastic energy. The CPN seeks elastic energy-minimizing position and orientation. Fig. 2b shows trajectories of CPN approaching SP from different directions that we plot atop of the dipolar configuration of $n(r)$ around SP (blue lines) calculated using an ansatz.[3,4,15–17]

During its motion (Fig. 2a), CPN is free to rotate around $n_0$ as it is attracted to SP starting from different initial positions and orientations of a CPN–SP center-to-center separation vector $r_{cc}$. By rotating around $n_0$, the dipole and quadrupole moments induced by CPN also align with respect to the microsphere to further minimize the total elastic energy. In the SP vicinity with dipolar elastic distortions, $n(r)$ around the CPN can be modified as compared to that in a uniformly aligned LC (Fig. 1a) or can even transition between two states with different alignments of $p$ shown in the insets of Fig. 1a. However, the details of this behaviour cannot be probed by optical imaging due to spatial resolution limitations. After approaching SP from any direction, CPN slides along the sphere's surface (trajectories 1 and 4) to the equatorial plane (dashed black circles) where it elastically binds to SP (Fig. 2b) in the minimum-energy location and orientation. Equilibrium arrangement of CPN and SP is shown in Fig. 2a, b and e. The CPN resides in the equatorial plane (Fig. 2b and e) and can drift around SP unless restricted by the cell confinement. Fig. 2d and e show a schematic diagram of $n(r)$ around these interacting colloids at the onset and in the end of their interaction. In the final elastically bound state, $n(r)$ distortions due to CPN match distortions due to SP near its equator, which is reminiscent to dipole–quadrupole interactions between differently surface-treated colloidal microspheres.[7,8] The total elastic energy is likely further minimized via alignment of the elastic dipole moment of CPN to be anti-parallel with respect to that of SP (Fig. 2e and f).

We have measured the strength of elastic pair interactions by tracking the position[31,32] and orientation of particles with the dark-field video microscopy[21] (Fig. 2a) after SP was released from the laser trap. Since inertia effects are negligible due to the low Reynolds number[2] (Re << 1), the elastic interaction force is balanced by the Stokes drag force $F_S = c_d v(t)$, where cd is the average drag coefficient for a nanoparticle in the nematic determined using their experimentally measured self-diffusion coefficients.[31] Time-dependent center-to-center separation $r_{cc}(t)$ between two colloids (inset of Fig. 2c) allows us to determine the relative velocity $v(t) = dr_{cc}/dt$ of CPN with respect to the center of SP (Fig. 2a and b) and, consecutively, the separation-dependent elastic interaction force and potential (Fig. 2c). Experimental data (Fig. 2c) were fitted with the expression for the interaction potential in the form $U_a(r_{cc}) \propto a/r_{cc}^3 + bl_{CPN}/r_{cc}^4$ ($l_{CPN}$ is a length of the CPN nanoparticle and $a$, $b$ are fitting constants) taking into account elastic pair interactions between the SP dipole and the CPN colloid with elastic structure having both weak dipolar and strong quadrupolar moments.[19,20] Data fitting (Fig. 2c) yields $b/a \approx 10$, which indicates that the dipole–quadrupole[7] term prevails in elastic pair interactions between CPN and SP. This is a natural result considering that the very existence of the dipole moment is caused by the pentagonal shape of the CPN cross-section, a relatively small departure from the cylindrical shape for which only a quadrupole moment would be expected. The attractive elastic force measured at $r_{cc} \approx 2.5$ µm is within the range of 0.5–1 pN; the strongest attraction is observed when the nanoparticle approaches from the directions (see trajectories 3–6 in Fig. 2b) opposite to the hedgehog. Elastic binding of CPN takes place within an experimentally determined belt surrounding SP along its equator (Fig. 2f). The character of elastic interactions between SP and CPN is dependent on the orientation of the elastic dipole moment of CPN with respect to the SP elastic dipole, yielding the strongest attraction when they are anti-parallel (Fig. 2e and f). The combination of the effects of the CPN-excluded volume of elastic distortions, dissimilar particle sizes, and CPN's rotation around $n(r)$ yields the unexpected omnidirectional CPN–SP attraction (Fig. 2b).

**Elastic alignment and torque transfer to nanoprisms**

Attractive interactions between CPNs and SPs in a homeotropic cell (Fig. 3 and 4) are accompanied by a long-range torque transfer mediated by nematic elasticity (Fig. 3). At $r_{cc} >$

$10R_{SP} \gg d$ (Fig. 3a), two particles drift randomly due to the Brownian motion and CPN freely rotates to arbitrary azimuthal angles $\beta$ (Fig. 3a and c) having no correlation with the angle $\theta$ describing azimuthal orientation of $r_{cc}$ (Fig. 3a). However, at $r_{cc} \approx d$, CPN aligns along $r_{cc}$ so that $\beta = \theta$ (Fig. 3a–c). Once this "communication" between CPN and SP via transfer of the elastic torque is established, it is possible to quite precisely ($|\theta - \beta| < 5°$) control the orientation of CPN by changing the position of SP using optical tweezers (Fig. 3a–c). The alignment can be understood as the elastic torque due to director deformations created by SP. Fig. 3d shows that $n(r)$ deformations caused by SP propagate to distances larger than $d$. The CPN introduces additional distortions into $n(r)$ due to the SP dipole. The elastic energy is at minimum when the CPN's long axis orients along $r_{cc}$, giving rise to the elastic torque causing rotation of CPN if its orientation is different (Fig. 3). This elastic torque exerted on CPN is balanced by a hydrodynamic drag torque as $\Gamma_{el} = -c_\beta d(\theta - \beta)/dt$, where $c_\beta \approx 1.43 \times 10^{-19}$ N m s is the rotational drag coefficient estimated from the rotational diffusion coefficient determined via tracking fluctuations of $\beta$. The angular drift velocity $d(\theta - \beta)/dt$ depends on $r_{cc}$ and can be determined using the angular distribution shown in Fig. 3c. $\Gamma_{el}$ decreases when $r_{cc}$ increases being $\approx 0.06$ pN µm rad$^{-1}$ at $r_{cc} \approx 9$ µm and $\approx 0.4$ pN µm rad$^{-1}$ at $r_{cc} \approx 3$ µm (Fig. 4a and c). The observed effect of long-range elastic torque transfer can be used to control orientation of elongated nanoparticles. The attractive interactions between SPs and CPNs self-aligned along $r_{cc}$ overcome the Brownian motion and become well pronounced at $r_{cc} < 8R_{SP}$ (Fig. 4a). The dipole–quadrupole term dominates in the attractive potential $U_a$ and experimental data yield elastic binding energies of hundreds of $k_B T$ (Fig. 4b), an order of magnitude lower compared to those for microparticles.[22,27,31]

**Elastic pair interactions between colloidal microspheres and starfruit-shaped nanoprisms**

Elastic interactions between CSNs and SPs are strongly anisotropic (Fig. 5): they attract when $r_{cc} \| n_0$ (Fig. 5a, b and g–i) and repel when $r_{cc} \perp n_0$ (Fig. 5k–n). When CSN approaches SP from the side of the hedgehog and $r_{cc} \| n_0$, CSN attracts to the region near the hedgehog (Fig. 5a–c). Elastic binding between particles is weak, yielding equilibrium distance $r_{cc} \approx 3.5$ µm (Fig. 5d) and displacement distribution of width $\Delta r_{cc}$ ~700 to 1000 nm (Fig. 5d). For this configuration, the equilibrium state corresponds to the angle $\alpha \approx 40°$ between $r_{cc}$ and $n_0$ and an angle $\phi \approx 0°$ between the CSN's long axis and $n_0$ (Fig. 5e and f); $\alpha$ changes within ~20° to 45° (Fig. 5e and o). The angle distribution in Fig. 5f shows that CSN changes its orientation up to $\sim\pm 20°$ with respect to $n_0$ while adjusting to the local deformations of $n(r)$ (Fig. 5b and c). The CSN quadrupole is also strongly attracted to the SP dipole when positioned at $r_{cc}\|n_0$ from the side of the SP hemisphere opposite to the location of the hedgehog (Fig. 5g and h). The CSN quadrupole and SP dipole repel when positioned at about $45° < |\alpha| < 135°$ (Fig. 5e and o) with the maximum repulsion at $r_{cc}\perp n_0$ (Fig. 5k–n).

Elastic attraction and repulsion between the CSN quadrupole and the SP dipole can be qualitatively understood considering $n(r)$ around both particles (Fig. 5b, i, l and n). When CSN approaches the SP along $n_0$, it matches the local $n(r)$ of the elastic dipole (Fig. 5b and i). When $r_{cc}\perp n_0$, antagonistic orientation of $n(r)$ at the surfaces of two particles, normal at SP and planar at CSN (Fig. 5l and n), causes repulsion (Fig. 5f). The angular zones of attraction are shown by a green colour in Fig. 5o and were determined using videomicroscopy observations (Fig. 5a, g, k and m). SP–CSN colloidal pairs can be moved and arranged into different structures. Fig. 5j shows the arrangement of two SP–CSN pairs into a chain due to the elastic dipole–dipole interactions between two SPs. For example, one can build the chain of unidirectionally aligned gold nanoprisms.

## Conclusions

We have demonstrated that interactions between colloidal particles having dissimilar shapes, compositions, and dimensions can exhibit behaviour, such as omnidirectional attraction, not encountered for monodisperse colloids in LCs. Convex pentagonal nanoprisms align perpendicular to the far-field director and induce director distortions having both elastic dipole and quadrupole moments while concave nanoprisms align along the far-field director and induce elastic quadrupoles. Both nanoparticles elastically bind in well-defined positions and with well-defined orientations next to colloidal microspheres, with elastic binding energies on the order of hundreds of $k_BT$, allowing for the formation of well-defined colloidal pairs and other superstructures formed from them. Our findings pose new challenges for theoretical modelling of colloidal self-assembly in LCs and are of interest for applications in nanophotonics and nanoscale energy conversion.

**Acknowledgements.** This work was supported by the International Institute for Complex Adaptive Matter and the NSF grants no. DMR-0645461, DMR-0820579 and DMR-0847782. We thank Christian Schoen and Shelley Coldiron from Nanopartz, Inc. for providing nanoparticles. We acknowledge discussions with Taewoo Lee, Angel Martinez, Rahul Trivedi, and Qingkun Liu.

# Figures

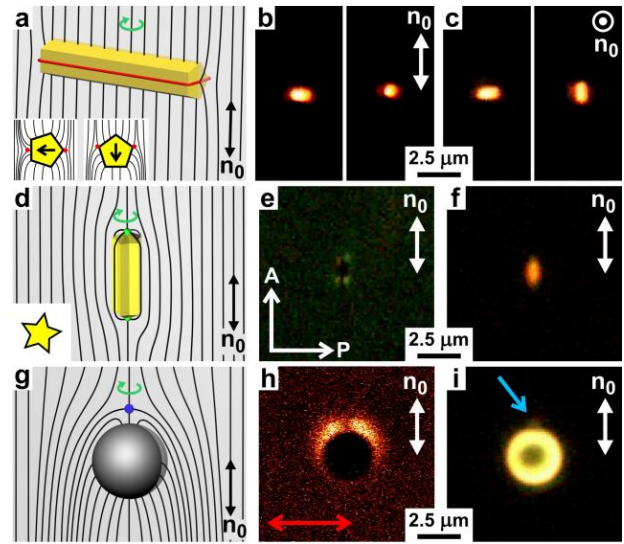

**Fig. 1** Colloidal particles in a nematic LC: (a, d and g) schematic diagrams of $n(r)$ (black lines) around CPN, CSN and SP, respectively; dark-field images of CPN in planar (b) and homeotropic (c) cells; polarizing (e) and dark-field (f) images of CSN in the planar cell; three-photon excitation fluorescence polarizing (h) and dark-field (i) images obtained in the planar cell. Red line in (a), green semispheres in (d), and blue sphere in (g) show "Saturn ring", boojums and hedgehog defects, respectively. Insets in (a and d) show the base shapes of CPN and CSN: black arrows in (a) show an elastic dipole moment of elastic dipoles induced by asymmetric $n(r)$ distortions around CPN; red dots indicate intersections of half-integer disclination. "A" and "P" in (e) mark the crossed polarizer and analyzer. Double white and black arrows show the far-field director $n_0$. Double red arrow in (h) shows the direction of polarization of the excitation light. Green arrows in (a, d and g) show allowed rotations of particles. Blue arrow in (i) points to a hedgehog.

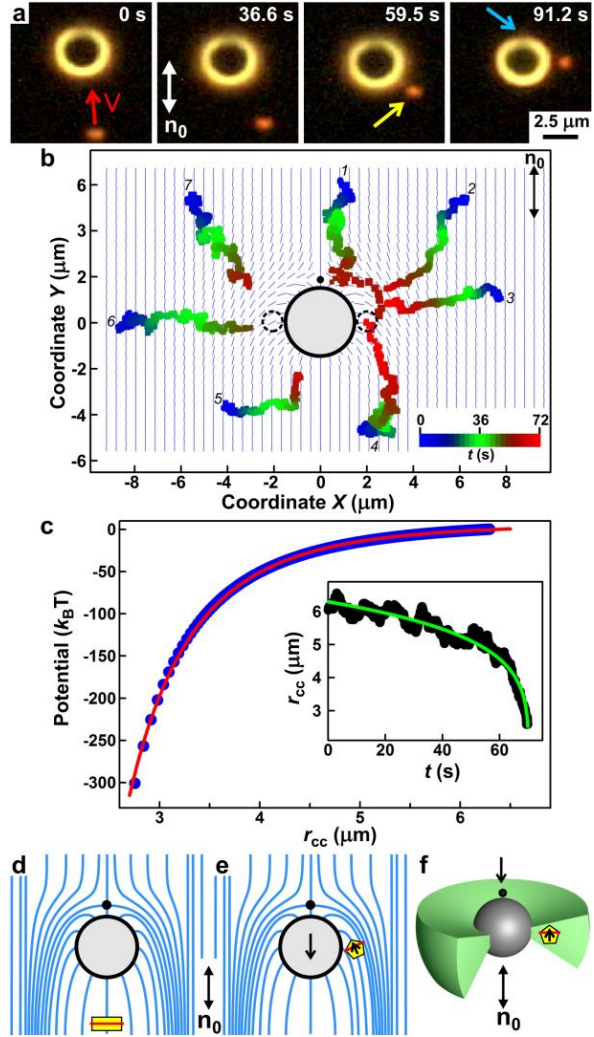

**Fig. 2** Elastic interactions between CPN and SP in a planar cell of $d \approx 10$ μm. (a) A sequence of dark-field images showing attractive interaction. (b) Superimposed trajectories of CPN in the vicinity of SP. (c) Attractive potential as a function of the center-to-center distance $r_{cc}$ between CPN and SP. The inset in (c) shows $r_{cc}$ vs. time. (d and e) $\boldsymbol{n}(\boldsymbol{r})$ around the SP with CNP in different locations. (f) A schematic diagram of the CNP binding belt (green) around the SP elastic dipole. Yellow and red arrows in (a) show CPN and the direction of its motion, respectively. Blue arrow in (a) and black filled circle in (b, d–f) show the hedgehog. Blue lines in (b, d and e) show $\boldsymbol{n}(\boldsymbol{r})$. Black arrows in (e and f) show the antiparallel elastic dipole moments of the structures induced around SP and CNP.

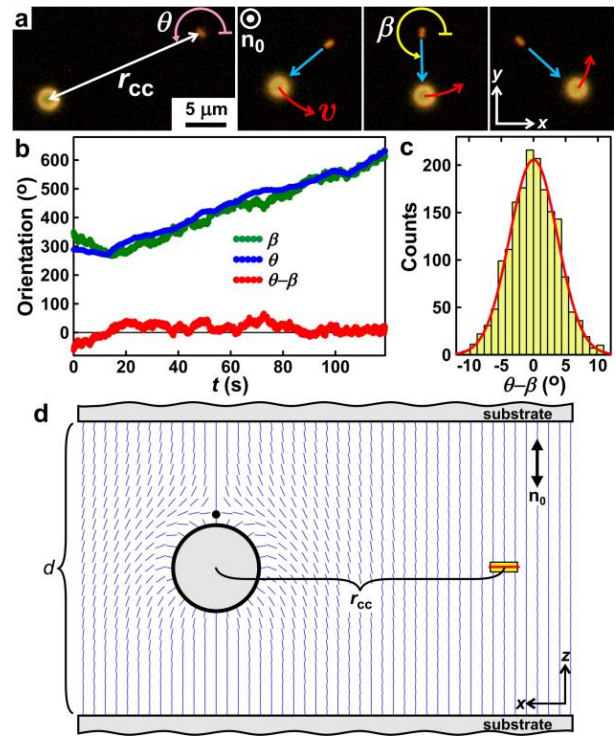

**Fig. 3** Elastic alignment of CPN by distortions due to the SP dipole in a homeotropic cell of $d \approx 10$ μm: (a) dark-field images showing orientational interaction between CPN and SP; starting from $r_{cc} \approx 10$ μm, the long axis of CPN is always pointing toward the center of SP (blue arrows); (b) orientation of the CPN's long axis vs. orientation of $r_{cc}$; (c) fluctuations of $\beta$ at equilibrium $r_{cc}$; (d) schematic of $n(r)$ deformations (blue lines) around SP and CPN.

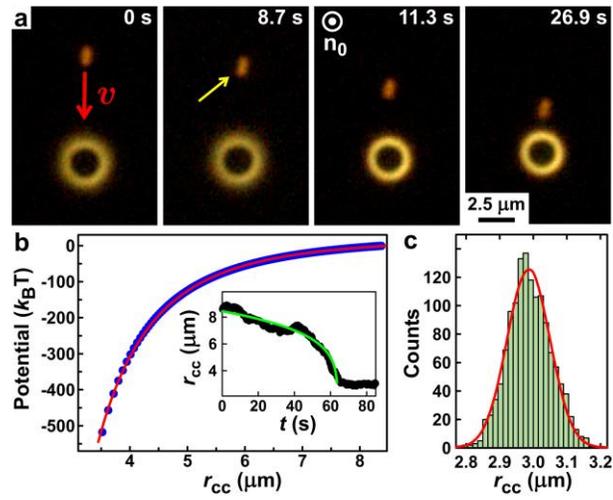

**Fig. 4** Alignment and attraction of CPN by the SP dipole in a homeotropic cell of $d\approx10$ μm: (a) dark-field images showing attraction between CPN and SP; (b) attractive potential as a function of the separation $r_{cc}$ between particles; (c) distribution of $r_{cc}$ for elastically bound CPN and SP in an equilibrium arrangement.

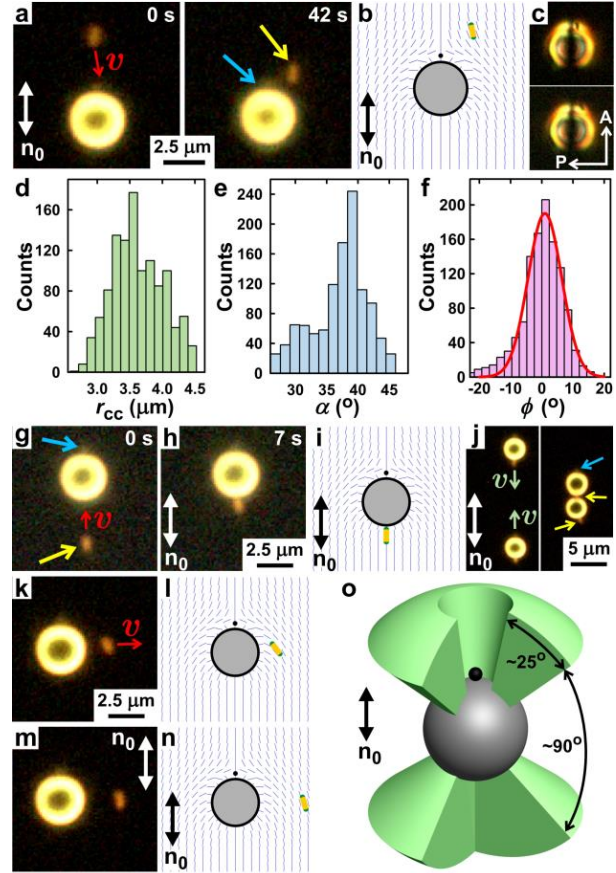

**Fig. 5** Elastic interactions between CSN and SP in a planar cell of $d \approx 10$ μm: (a, g, h, k and m) dark-field images showing the interaction (red arrow) between CSN and SP at different angles with respect to $n_0$; (b, i, l and n) schematics of the corresponding $n(r)$ (blue lines); (c) polarizing textures of CSN elastically trapped in the defect region of SP; (d) distribution of $r_{cc}$ between attracted particles in (c); (e) distribution of the orientation of $r_{cc}$; (f) distribution of the orientation of the CSN's long axis with respect to $n_0$; (j) dark-field images of the attraction (green arrows) between elastic dipoles with the elastically bound CSN; (o) a schematic showing zones of attraction (green) between CSN and SP. Yellow rectangles with green semispheres at their ends show CSNs with boojums.